\documentclass[conference]{IEEEtran}
\IEEEoverridecommandlockouts

\usepackage{cite}
\usepackage{amsmath,amssymb,amsfonts}
\usepackage{algorithmic}
\usepackage{graphicx}
\usepackage{textcomp}
\usepackage{xcolor}
\usepackage{amsmath,amsthm,amssymb,paralist,subfigure,cite,graphicx,color}
\usepackage{algorithm2e}
\usepackage{cuted,mathtools,lipsum}
\usepackage{multicol,graphicx}
\usepackage{multirow}
\usepackage{lipsum,booktabs}
\usepackage{fancyhdr}

\fancypagestyle{firstpage}{
    \fancyhf{}
    \fancyhead[C]{\footnotesize Proceedings of the 12th RSI International Conference on Robotics and Mechatronics (ICRoM 2024), Dec. 17-19, 2024, Tehran, Iran}
    \fancyfoot[L]{\scriptsize 979-8-3315-2973-4/24/\$31.00 \textcopyright 2024 IEEE}
}

\def\mb{\mathbf}
\def\mc{\mathcal}

\def\BibTeX{{\rm B\kern-.05em{\sc i\kern-.025em b}\kern-.08em
    T\kern-.1667em\lower.7ex\hbox{E}\kern-.125emX}}
\begin{document}

\title{Decentralized Mobile Target Tracking Using Consensus-Based Estimation with Nearly-Constant-Velocity Modeling}

\author{\IEEEauthorblockN{ Amir Ahmad Ghods}
\IEEEauthorblockA{\textit{Faculty of Mechanical Engineering,} \\
\textit{Semnan University,}
Iran. \\
amir-ghods@semnan.ac.ir}
\and
\IEEEauthorblockN{Mohammadreza Doostmohammadian}
\IEEEauthorblockA{\textit{Mechatronics Department,} \\
\textit{Faculty of Mechanical Engineering, }\\
\textit{Semnan University,}
Iran. \\
doost@semnan.ac.ir}
}

\maketitle
\thispagestyle{firstpage}

\begin{abstract}
Mobile target tracking is crucial in various applications such as surveillance and autonomous navigation. This study presents a decentralized tracking framework utilizing a Consensus-Based Estimation Filter (CBEF) integrated with the Nearly-Constant-Velocity (NCV) model to predict a moving target's state. The framework facilitates agents in a network to collaboratively estimate the target's position by sharing local observations and achieving consensus despite communication constraints and measurement noise. A saturation-based filtering technique is employed to enhance robustness by mitigating the impact of noisy sensor data. Simulation results demonstrate that the proposed method effectively reduces the Mean Squared Estimation Error (MSEE) over time, indicating improved estimation accuracy and reliability. The findings underscore the effectiveness of the CBEF in decentralized environments, highlighting its scalability and resilience in the presence of uncertainties.

\end{abstract}


\section{Introduction}
\IEEEPARstart{M}{obile} target tracking is a critical task in numerous applications, including surveillance, autonomous navigation, and defense systems. Accurate real-time tracking of moving objects requires the development of advanced algorithms that integrate sensor data, predict target trajectories, and compensate for uncertainties arising from environmental noise. Commonly used dynamic models, such as the Nearly-Constant-Velocity (NCV) \cite{c1,c2} and Nearly-Constant-Acceleration (NCA) models \cite{c3}, are favored for their simplicity and effectiveness in approximating target motion.

In recent years, decentralized consensus algorithms have emerged as a promising approach in multi-agent systems for target tracking \cite{c4,c5,c6,c7,c8,c9}. In contrast to centralized methods \cite{c10,c11}, where a single node aggregates information and makes decisions, decentralized algorithms distribute the computational load across multiple agents, enhancing system scalability, resilience, and robustness against communication failures. These algorithms enable agents to collaboratively estimate the target’s state by sharing local observations and fusing information, achieving consensus on the target’s position without needing a central coordinator.

Despite their advantages, decentralized consensus algorithms encounter significant challenges, such as varying communication delays, network topology design constraints, limited bandwidth, and asynchronous updates, all of which can adversely affect tracking performance \cite{c110,c111,c112,c250,c251}. Another critical issue is the presence of noisy sensor data, which can arise from various factors such as environmental interference, electrical noise, and quantization errors. To address these challenges, the use of filtering techniques has been widely proposed in the signal processing literature as an effective means of mitigating the impact of noise on system performance.

In decentralized target tracking, various filtering techniques are employed to estimate and monitor the state of a target across a distributed network of sensors or agents. One widely utilized approach is the Kalman filter, which is effective for estimating the state of a dynamic system from a series of noisy measurements. The Kalman filter leverages the system's motion model to predict the next state based on the current estimate and updates the covariance matrix to reflect the uncertainty in the prediction.

Several variations of the Kalman filter have been developed to address specific challenges in decentralized tracking. The Extended Kalman Filter (EKF) is an adaptation designed for nonlinear systems\cite{c12,c13,c14}. It linearizes the nonlinear system dynamics and measurement models around the current estimate and is commonly applied in multi-agent systems where the target dynamics or measurement models exhibit nonlinearity. Each agent in such systems performs the EKF locally and shares its results with neighboring agents to enhance the global estimation.

Another notable variation is the Unscented Kalman Filter (UKF), which employs a set of sigma points to approximate the probability distribution of the state\cite{c15,c16,c17,c18}. The UKF offers greater accuracy than the EKF in handling highly nonlinear systems. It is particularly useful when target dynamics or sensor models are significantly nonlinear, with each agent executing the UKF and collaborating with others to refine the estimate.

A more recent development is the Distributed Kalman Filter (DKF), which extends the Kalman filter to decentralized environments \cite{c20,c21,c22}. In this approach, each agent maintains its own Kalman filter and exchanges local estimates with neighboring agents to collaboratively refine the global state estimate. This method is suitable for systems where agents need to process and exchange state estimates to reach a consensus on the target state.

The Consensus Kalman Filter (CKF) combines Kalman filtering with consensus algorithms, allowing agents to iteratively update their state estimates based on shared information. This approach is widely used in decentralized networks where agents must achieve consensus on the state estimate and are particularly effective in scenarios characterized by noisy measurements and communication constraints\cite{c19,c21,c22,c23,c24,c25}.

Another example of filtering in distributed target tracking is the Consensus-Based Estimation (CBE) filter. This widely-used distributed technique in sensor networks and multi-agent systems allows agents or sensors to collaboratively estimate the state of a system. By exchanging information with neighboring agents, the system seeks to reach a consensus on the estimated state\cite{c26,c27,c28,c29,c30,c31,c32,c33,c34,c35,c36}. A variant of this filtration technique, the Saturation-Based fault mitigation filter, can be structured as a Consensus-Based Estimation filter with integrated fault tolerance, enhancing the robustness of decentralized target tracking systems. This approach incorporates Consensus-Based filtering principles with a saturation mechanism to mitigate the effects of faulty or outlier measurements \cite{c37,c38,c39,c40}. In this framework, multiple agents estimate the state of a mobile target by processing their local measurements and exchanging information with neighboring agents. The consensus algorithm aims to achieve convergence among agents on a unified estimate of the target’s state, despite potential differences in individual measurements.

In this work, we propose a novel approach to decentralized mobile target tracking by assuming that the target moves with an approximately constant velocity over short time intervals. This assumption justifies the use of the Nearly-Constant-Velocity (NCV) model for the target's dynamics, a commonly used framework in literature. However, unlike previous works, we demonstrate that adopting the NCV model in our system not only improves computational efficiency but also reduces simulation costs, making it particularly advantageous for real-time, large-scale applications.

Furthermore, we enhance localization accuracy by incorporating a customized variant of the multilateration technique based on the Time-of-Arrival (TOA) of beacon signals transmitted by the target and measured by distributed agents. Our approach, building upon \cite{c4}, introduces a novel method for calculating the observation matrix that optimally relates the system’s internal states to the measurements. This approach significantly reduces communication overhead and decision complexity, thereby improving network scalability which is a key challenge in large communication networks.

To address the common issue of noise and measurement variations from agents, we introduce an innovative saturation-based filtering scheme, expanding on \cite{c37}. Unlike traditional filtering techniques, our approach effectively mitigates measurement errors while maintaining robustness in diverse environmental conditions. These collective innovations provide a significant improvement over existing tracking systems, enhancing both the efficiency and scalability of decentralized networks, and making our method highly suitable for large-scale, real-time tracking applications.

The remainder of this paper is organized as follows: Section \ref{graph} provides a brief review of graph theory, which is fundamental for understanding the subsequent sections. In Section \ref{sysmodel}, we present the system model and problem formulation, divided into two subsections: Target Dynamic Model Using the Nearly-Constant-Velocity Approach and Measurement Model for Agents and Observation Matrix Derivation. Section \ref{filt} introduces the Consensus-Based Estimation Filter, followed by the simulation process and results in Section \ref{sim}. Finally, Section \ref{conclusion} concludes the paper.

\begin{table}
	\caption{Description of Variables used in this Study}
	\setlength{\tabcolsep}{0.7\tabcolsep}
	\centering
	\begin{tabular}{ *{2}{c} }
		\toprule
		\textbf{Variable} & \textbf{Description} \\
		\midrule
		$\mc{G}$ & A communication network of agents.   \\ 
		$\mc{N}$ & Set of agents in the network.   \\ 
		$\mc{E}$ & Set of links representing connections between agents. \\
		$\mc{W}$ & Weight matrix that represents attributes of connections.\\
		$\mc{N}_i$ &Set of neighboring nodes directly connected to node $i$.   \\
		$|\mc{N}_i|$ & Number of neighboring nodes directly connected to node $i$.   \\
		$\textbf{A}$ & Transition matrix.           \\
		$\textbf{B}$ & Input matrix.\\
		$\mb{x}$ & State vector.   \\
		$\Delta$ & Sampling time.\\
		$k$ & Iteration.   \\
		$\mb{w}$ &  Process noise.           \\
		$\textbf{H}$ & Observation matrix\\
		$\mb{v}$ & Observation noise.   \\
		$\mb{y}$ & Observed measurement.           \\
		${\xi}$ & A scalar representing observation confidence. \\
		$g$ & A gain for saturating the estimation innovation.   \\
		$L$ & Communication rate.           \\
		\bottomrule
	\end{tabular}
\end{table}

\section{a Brief Review of Graph Theory}\label{graph}
In the context of graph theory, a network of agents can be formally represented as  $\mc{G}=\{\mc{N},\mc{E},\mc{W}\}$, where $\mc{N}=\{1,\dots,n\}$ denotes the set of agents (nodes), $\mc{E}=\{(i,j)|i \rightarrow j,~i,j \in \mc{N}\}$ represents the set of links (edges) between them, and $\mc{W}=[\mc{W}_{ij}]$ is the weight matrix associated with these connections. The neighboring set, $\mc{N}_i$ identifies the nodes that are directly connected to a given node $i$. A graph is considered undirected when the edges between nodes lack a specific direction, whereas a graph is classified as directed if the edges indicate a direction from one node to another. The weight matrix $\mc{W}$ encodes various attributes of the connections, such as distance, cost, or time. For the purposes of this paper, we will utilize an undirected graph to highlight the mutual interactions between agents. Furthermore, we assume the weight matrix to be row-stochastic($\sum_{j=1}^n \mc{W}_{ij}=1$), meaning that each row consists of non-negative entries that sum to 1, ensuring a consistent probabilistic interpretation of the connection strengths.

\section{System Model and Problem Formulation}\label{sysmodel}
This section presents the dynamic model for mobile targets, followed by the implementation of the Nearly-Constant-Velocity (NCV) model, as proposed in \cite{c4}, to approximate the target's motion. The measurement model for the multi-agent system is then introduced, and the corresponding observation matrix is derived.
\subsection{Target Dynamic Model Using the Nearly-Constant-Velocity Approach}
A dynamic model of a target system offers a mathematical framework for describing the temporal evolution of the target's dynamics. This evolution is captured by the following equation:

\begin{equation}  \label{eq_targ}
\mb{\mb{x}}{(k+1)} = \textbf{A}\mb{x}{(k)}+\textbf{B}\mb{w}{(k)},
\end{equation}

In this equation, $\mb{x}{(k)}$ denotes the current state of the system, while $\mb{x}{(k+1)}$ represents the predicted state at the next time step. The matrix $\textbf{A}$, known as the transition matrix, characterizes the probabilities of transitioning from one state to another over discrete time intervals. The matrix $\textbf{B}$, referred to as the input matrix, details how external inputs or control signals influence the system's state. Additionally, $\mb{w}{(k)}$ represents the process noise, encompassing random variations or disturbances that impact the state of the system but are not directly controlled or accounted for within the system's model. This process noise captures the uncertainties or imperfections inherent in the system's behavior and modeling.

In target tracking, the Nearly-Constant-Velocity (NCV) model offers an efficient framework for predicting a target's future position and velocity by assuming that velocity remains approximately constant over short intervals. This method simplifies the tracking process and improves computational efficiency, making it particularly effective in scenarios with minimal velocity variations. The NCV model can be expressed within the dynamic modeling framework as follows, with parameters substituted accordingly:

\small
\begin{equation}  
\textbf{A} = \left(
\begin{array}{cccccc}
    1 & 0 & 0 & \Delta & 0 & 0\\
    0 & 1 & 0 & 0 & \Delta & 0  \\ 
    0 & 0 & 1 & 0 & 0 & \Delta \\ 
    0 & 0 & 0 & 1 & 0 & 0 \\ 
    0 & 0 & 0 & 0 & 1 & 0 \\ 
    0 & 0 & 0 & 0 & 0 & 1 \\    
\end{array} \right)
,~ 
\textbf{B} = \left(
\begin{array}{ccc}
    \frac{\Delta^2}{2} & 0 & 0\\
    0 & \frac{\Delta^2}{2} & 0  \\ 
    0 & 0 & \frac{\Delta^2}{2} \\ 
    \Delta & 0 & 0 \\ 
    0 & \Delta & 0 \\ 
    0 & 0 & \Delta \\    
\end{array} \right). 
\label{eq_G_ncv}
\end{equation}
\normalsize
Where $\Delta$ represents the sampling time, the proposed matrix \textbf{A} is implemented as the transition matrix, while matrix \textbf{B} serves as the input matrix to incorporate acceleration as process noise in the system.

With the appropriate application of these parameters, the state vector $\mb{x}$ can be expressed as $\mb{x}=(p_x, p_y, p_z, v_x, v_y, v_z)^T$ .Notably, in each iteration, both velocity and position components are obtained. 

\subsection{Measurement Model for Agents and Observation Matrix Derivation}
The measurement model reflects the data acquired from individual agents at each iteration and is formulated as follows:

\begin{equation}  \label{eq_agents}
\mb{\mb{y}}_{i}{(k)} = \textbf{H}_{i}\mb{x}{(k)}+\mb{v}_i{(k)},
\end{equation}

In Equation \eqref{eq_agents}, $\mb{\mb{y}}_{i}{(k)}$ denotes the observed measurement obtained from each agent at iteration $k$, while $\textbf{H}_{i}$ represents the observation matrix for each agent that relates the system’s internal state to the corresponding measurements. The term $\mb{v}_i{(k)}$ refers to the observation noise, capturing the uncertainties or errors inherent in the measurements acquired from the system.\\
\begin{figure}
    \centering
    \includegraphics[width =0.5\textwidth]{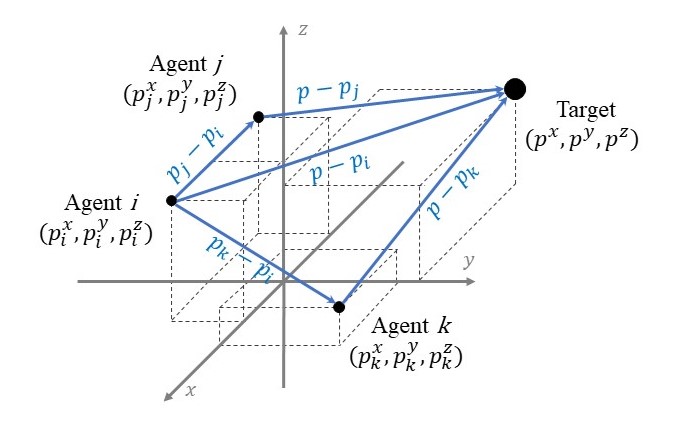}
    \caption{Illustration of the derivation of the observation matrix for a system with three agents, based on the proposed multilateration technique.}
    \label{fig11}
\end{figure}
To derive the observation matrix, we utilize a multilateration technique based on Time-of-Arrival (TOA), as described in \cite{c4}. In this method, a beacon signal transmitted by the target at a known speed is employed for localization, as shown in Fig.~\ref{fig11}. Since each agent only communicates with its immediate neighbors, the process is simplified by linearizing the observation matrix. The resulting linearized observation matrix is expressed as follows:

\begin{equation}
\begin{cases}
	 {p}^x_{j,i} =  p^x_{j} - p^x_{i}\\
	 {p}^y_{j,i} = p^y_{j} - p^y_{i}\\
	 {p}^z_{j,i} = p^z_{j} - p^z_{i} \\
\end{cases}
\textbf{H}_{i} = \left(
\begin{array}{cc} \label{eq_om}
	\mb{p}_{j_1,i}^\top & \mb{0}_3^\top \\ 
	\vdots & \vdots \\
    \mb{p}_{j_{|\mc{N}_i|},i}^\top & \mb{0}_3^\top
\end{array} \right). 
\end{equation}

In equation \eqref{eq_om}, $p_{n,i} = ({p}^x_{n,i}, {p}^y_{n,i}, {p}^z_{n,i})$ denotes the relative positions of the neighboring agent $n$ with respect to agent $i$.

\section{Consensus-Based Estimation Filter }\label{filt}
In the field of target tracking, the application of estimation filters is crucial for enhancing system performance, particularly by mitigating the effects of noise and signal variations in real-time measurements. Filters not only improve the accuracy of these measurements but also stabilize control responses and provide reliable state estimation, even when data is noisy or incomplete.

Various studies have explored the effectiveness of different filter designs. A notable approach, as presented in \cite{c37}, employs a Saturation-Based mechanism to handle noise and faulty data. This filter, described by the following equation, provides an estimation of the target's state:

\begin{equation} \label{eq_fil}
{\hat{x}_i}{(k)}={A}{\hat{x}_i}{(k-1)}+g_i(k){{H}_{i}^T({{y}}_{i}{(k)}-{H}_{i}{A}{\hat{x}_i}{(k-1)})}
\end{equation}

Here, ${\hat{x}_i}{(k)}$ represents the estimated state of agent $i$ at time step $k$, $A$ is the state transition matrix, $H_i$ is the measurement matrix, and ${y_i(k)}$ is the observed measurement. The term $g_i(k)$, defined in Equation \eqref{eq_sat}, is a gain that regulates the filter response, particularly in the presence of noise:

\begin{equation} \label{eq_sat}
g_i(k)=
\begin{cases}
	 1, if {|{{y}}_{i}{(k)}-{H}_{i}{A}{\hat{x}_i}{(k-1)}|} \leq {\xi}\\
	\frac{\xi}{|{{y}}_{i}{(k)}-{H}_{i}{A}{\hat{x}_i}{(k-1)}|}, otherwise.\\
	
\end{cases}
\end{equation}

In this formulation, $g_i(k)$ introduces saturation to the estimation innovation term, which is the difference between the actual and predicted measurements. This saturation helps to mitigate the effect of large deviations that may arise due to noise or faulty data. The parameter $\xi$, called the observation confidence parameter, is chosen by the designer to limit the influence of these deviations, thereby ensuring robustness.

A consensus mechanism is employed to further refine the filter's performance in multi-agent systems. In this approach, each agent exchanges information with its neighbors multiple times per iteration in order to collectively agree on the estimated state of the target. This process ensures that agents maintain consistent state estimates, even in the presence of communication delays or differing local measurements. The consensus update equation is given by:
\begin{equation} \label{eq_con}
{\hat{x}_{i,l}}{(k)}={\hat{x}_{i,l-1}}{(k)}-{\epsilon}{\sum_{J \in {\mc{N}_i}}}{({\hat{x}_{i,l-1}}{(k)}-{\hat{x}_{J,l-1}}{(k)})}
\end{equation}

In this equation, ${\hat{x}_{i,l}}{(k)}$ represents the state estimate of agent $i$ after $l=1, ... ,L$ communication steps at iteration $k$, while ${\epsilon}$ is a small positive constant that controls the rate of convergence. The set ${\mc{N}_i}$ represents the neighbors of agent $i$, and the term  ${\epsilon}{\sum_{J \in {\mc{N}_i}}}{({\hat{x}_{i,l-1}}{(k)}-{\hat{x}_{J,l-1}}{(k)})}$ ensures that the estimate of agent $i$ moves toward the average of its neighbors' estimates.

The parameter $L$, representing the number of communication steps between agents per iteration, is crucial in determining the speed of consensus in distributed estimation. A higher $L$ increases communication frequency, accelerating convergence to a common state estimate and improving accuracy, particularly in noisy environments. Conversely, a lower $L$ slows convergence, which can result in less accurate estimates. 

While increasing $L$ enhances filter performance by facilitating faster consensus and reducing local estimation errors, it also raises communication and computational costs, which may be a concern in resource-limited systems like sensor networks. Therefore, the choice of $L$ must balance the need for rapid convergence with available communication bandwidth.

\section{Simulation and Results}\label{sim}
In this section, we first summarize the formulation outlined in Sections \ref{sysmodel} and \ref{filt} in the form of pseudocode, represented in Algorithm \ref{alg}. We then simulate the proposed model using MATLAB, followed by a detailed discussion of the results.

\begin{figure}
    \centering
    \includegraphics[width =0.4\textwidth]{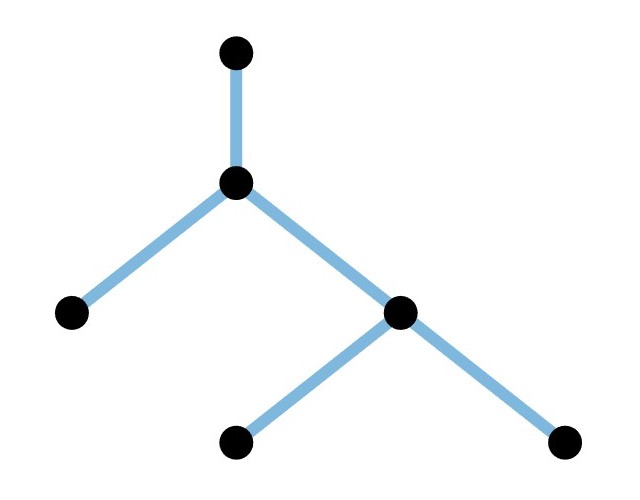}
    \caption{A random communication network consisting of six agents.}
    \label{fig1}
\end{figure}

In the simulation, we consider a communication network consisting of 6 agents, modeled as an Erdős–Rényi random graph (Fig.~\ref{fig1}). Algorithm \ref{alg}, which summarizes the formulation described in Sections \ref{sysmodel} and  \ref{filt}, is implemented. The algorithm first computes the observation matrix for each agent using Equation \eqref{eq_om}. In the Measurement Update step, Equations \eqref{eq_fil} and \eqref{eq_sat} are applied to derive the saturation-based filter, and consensus among the agents is subsequently achieved in the Estimate Consensus step using Equation \eqref{eq_con}.

For the simulation, the designer-specified parameters $\xi$ and $L$ are set to 1 and 10, respectively, with the weight matrix represented in a row-stochastic form.

\begin{figure}
\label{fig2}
    \centering
    \subfigure[]{
    \includegraphics[width=1\linewidth]{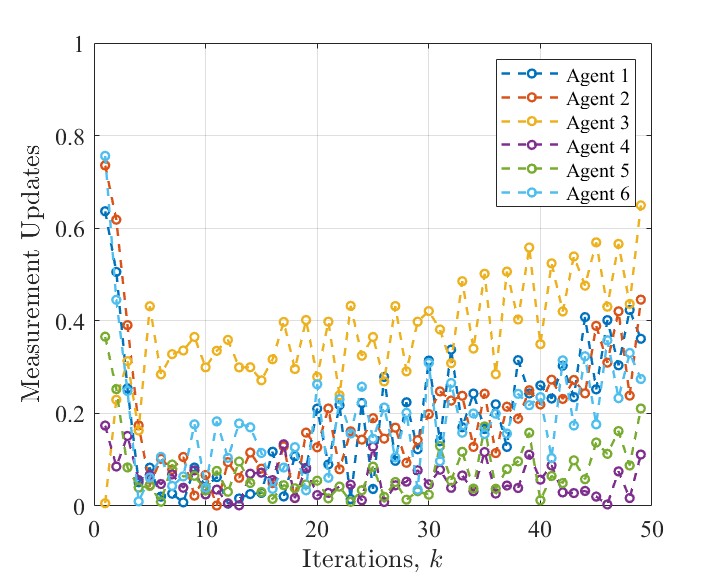} \label{fig2a}
    }
    \subfigure[]{
    \includegraphics[width=1\linewidth]{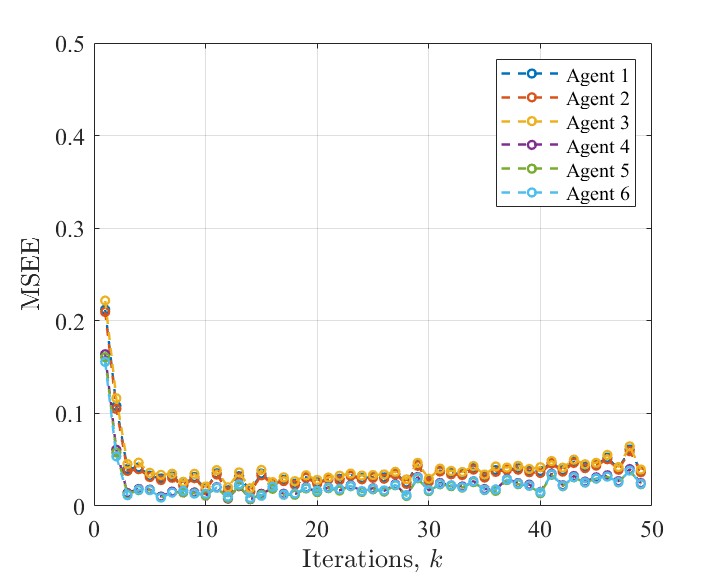}	\label{fig2b}
    }
    \subfigure[]{
    \includegraphics[width=1\linewidth]{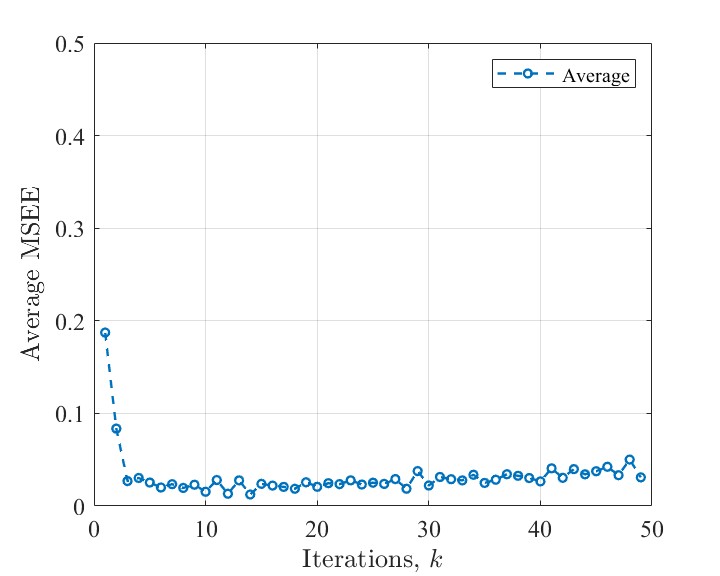}	\label{fig2c}
    }
    \caption{Results of the proposed algorithm for a communication network comprising 6 agents. (a) Measurement updates for each individual agent, (b) Mean Squared Estimation Error (MSEE) for each agent, (c) Average MSEE for the network.}
\end{figure}

The simulation results are illustrated in Figure 3. Fig.~\ref{fig2a} shows the measurement updates for each agent, represented as $\mb{m}=|{{y}}_{i}{(k)}-{H}_{i}{A}{\hat{x}_i}{(k-1)}|$. In Fig.~\ref{fig2b}, the Mean Squared Estimation Error (MSEE) between the agents' measurements and the moving target is presented, and computed as $e_{i} ={\frac{1}{k}}\sum_{k=1, 2, ...}{({\hat{x}}_{i}{(k)}-{x}(k))^2} $. Fig.~\ref{fig2c} illustrates the average MSEE, calculated as $e_{Avg} =	\frac{1}{|\mc{N}|}(\sum_{i=1}^{|\mc{N}|}{e_{i}})  $. 

An examination of Figs.~\ref{fig2b} and \ref{fig2c} reveals a decreasing trend in both the individual and average MSEE over time. This indicates that the network's collective estimation of the target’s position becomes increasingly accurate, demonstrating convergence of the proposed algorithm.

\RestyleAlgo{ruled}
\SetKwComment{Comment}{/* }{ */}
\begin{algorithm}
\caption{Decentralized mobile target tracking using the Consensus-Based Estimation Filter algorithm.}\label{alg}

\textbf{Initialize variables=} $({\hat{x}_i}(0), {\xi}, L, {\epsilon})$ \\
\For{$k=1, 2, ... $}{
    \For{$i=1, ..., \mc{N}  $}{
\textbf{Observation Matrix Calculation:}\\
\For{$j=1, ..., |\mc{N}_i|  $}{

	  $\textbf{H}_{i}$  {"Calculated from Equation (\ref{eq_om})"}
    }

\textbf{Measurement Update:}\\
$g_i(k)=min(1,\frac{\xi}{|{{y}}_{i}{(k)}-{H}_{i}{A}{\hat{x}_i}{(k-1)}|})$\\
${\hat{x}_i}{(k)}={A}{\hat{x}_i}{(k-1)}+g_i(k){{H}_{i}^T({{y}}_{i}{(k)}-{H}_{i}{A}{\hat{x}_i}{(k-1)})}$

\textbf{Estimate Consensus:} Let ${\hat{x}_{i,0}}{(k)}={\hat{x}_{i}}{(k)}$ \\
\For{$l=1, ..., L $}{
${\hat{x}_{i,l}}{(k)}={\hat{x}_{i,l-1}}{(k)}-{\epsilon}{\sum_{J \in {\mc{N}_i}}}{({\hat{x}_{i,l-1}}{(k)}-{\hat{x}_{J,l-1}}{(k)})}$
}
Let ${\hat{x}_{i}}{(k)}={\hat{x}_{i,L}}{(k)}$
}
}
\end{algorithm}

\section{Conclusions} \label{conclusion}
In this paper, we presented a decentralized target tracking framework using a Consensus-Based Estimation (CBE) filter tailored for mobile targets. The proposed model integrates the Nearly-Constant-Velocity (NCV) dynamic approach to predict the state of a moving target and incorporates a consensus mechanism among a network of agents. The dynamic model is implemented alongside a decentralized filtering technique, which allows agents to exchange local observations and iteratively update their estimates, achieving consensus on the target's state.

Through the use of the Saturation-Based filtering scheme, we addressed noise and variations in the agents' measurements, enhancing the robustness of the tracking process. The simulation results demonstrated that our approach achieves convergence, as reflected in the decreasing Mean Squared Estimation Error (MSEE) across agents over time, indicating accurate and reliable estimation. The consensus process, governed by the communication rate and local interactions, proved effective in ensuring agreement among agents, even in the presence of measurement noise and communication constraints.

\vspace{12pt}

\end{document}